\documentclass[oldversion]{aa}
\usepackage{graphicx}
\begin{document}
\headnote{Research Note}
   \title{Improved age constraints for the AB Dor quadruple system
\thanks{Based on observations collected at the European Southern Observatory, 
Chile (NACO SDI commissioning run, February 2004).}
}

   \subtitle{The binary nature of AB Dor B}

   \author{Markus Janson\inst{1} \and
          Wolfgang Brandner\inst{1,5} \and
          Rainer Lenzen\inst{1} \and
          Laird Close\inst{2} \and
          Eric Nielsen\inst{2} \and
          Markus Hartung\inst{3} \and
	  Thomas Henning\inst{1} \and
          Herv\'e Bouy\inst{4}
          }

   \offprints{Markus Janson}

   \institute{Max-Planck-Institut f\"ur Astronomie, K\"onigstuhl 17,
              D-69117 Heidelberg, Germany\\
              \email{janson@mpia.de, brandner@mpia.de, lenzen@mpia.de, henning@mpia.de}
         \and
             Steward Observatory, University of Arizona, 933 N.\ Cherry Ave, Tucson, AZ-85721-0065, USA\\
	\email{lclose@as.arizona.edu, enielsen@as.arizona.edu}
	\and
	    European Southern Observatory, Alonso de Cordova 3107, Santiago 19, Chile\\
	\email{mhartung@eso.org}
	\and
	    Astronomy Dpt. of UC Berkeley, 601 Campbell Hall, Berkeley, CA-94720-3411, USA\\
	\email{hbouy@astro.berkeley.edu}
	\and
	    UCLA, Div. of Astronomy, 475 Portola Plaza, Los Angeles, CA-90095-1547, USA\\
	\email{brandner@astro.ucla.edu}
             }

   \date{Received ---; accepted ---}

   \abstract{
We present resolved NACO photometry of the close binary AB Dor B in H- and Ks-band. AB Dor B is itself known to be a wide binary companion to AB Dor A, which in turn has a very low-mass close companion named AB Dor C. These four known components make up the young and dynamically interesting system AB Dor, which will likely become a benchmark system for calibrating theoretical pre-main sequence evolutionary mass tracks for low-mass stars. However, for this purpose the actual age has to be known, and this subject has been a matter of discussion in the recent scientific literature. We compare our resolved photometry of AB Dor Ba and Bb with theoretical and empirical isochrones in order to constrain the age of the system. This leads to an age estimate of about 50 to 100 Myr. We discuss the implications of such an age range for the case of AB Dor C, and compare with other results in the literature.
   
   }

   \maketitle
%
%________________________________________________________________

\section{Introduction}

AB Dor A is one of the most active late-type stars in the solar neighbourhood. Because of its short rotational period of approximately 0.514 days (Pakull 1981, Innis et al. 1998), resulting in a high level of activity and variability in X-rays (e.g.\ K\"urster et al. 1997, Schmitt et al. 1998, Sanz-Forcada et al. 2003), and at optical (e.g. Collier Cameron et al. 1999, Cutispoto et al. 2001) and radio wavelengths (e.g. Lim et al. 1994), it was initially classified as an evolved star, possibly a RS CVn or BY Dra type variable (e.g. Pakull 1981). Its high Lithium abundance, however, soon led to the suggestion that AB Dor is not an evolved star, but still in its post-T Tauri evolutionary phase (e.g. Rucinski
1983, Vilhu et al. 1987). Based on the Hipparcos parallax measurements for AB Dor (HIC 25647) of $\pi (\rm{abs})$ = 66.92$\pm$0.54 mas, corresponding to a distance of 14.94$\pm$0.12 pc (Perryman et al. 1997), Wichmann et al. (1998) concluded that AB Dor is a zero-age-main-sequence
star of spectral type K1.

AB Dor is listed in the Index catalogue of visual Double Stars (IDS, Jeffers et al. 1963) as a binary star with a separation of about 10$''$. The visual companion Rst 137B (Rossiter 1955) is of spectral type M5 (Mart\'{\i}n \& Brandner 1995). Radial velocity measurements (Innis et al. 1986) and relative
astrometry combined with proper motion measurements (Innis et al. 1986, Mart\'{\i}n \& Brandner 1995) confirm that Rst 137B and AB Dor form a physical binary, hence Rst 137B can be referred to as AB Dor B.

Due to the presence of AB Dor C, which is a close, low-mass companion to AB Dor A that was detected dynamically by Guirado et al. (1997), and recently resolved by Close et al. (2005), the AB Dor system is interesting in terms of calibrating theoretical models for young, low-mass objects. Since the mass of AB Dor C is known from the dynamical measurements, and the brightness is known from the resolved imaging, mass-luminosity relationships can be calibrated if the age of the system is known. This is however not a trivial issue, as is evident by the recent discussion in the scientific literature (see e.g. Close et al. 2005 -- hereafter C05, Luhman et al. 2005 -- hereafter L05, Nielsen et al. 2006, L\'opez-Santiago et al. 2006).

As suggested by Pakull (1981) and Rucinski (1985), precise age-dating of AB Dor B will yield a good determination of the evolutionary status of the AB Dor system as a whole. Here we present resolved NACO photometry of the two components of AB Dor B, allowing for placement of the individual objects in a color-magnitude diagram for comparison with theoretical and empirical isochrones. We use this to estimate an age range for the AB Dor system and apply it to AB Dor C. We compare our results to L05, in which a similar analysis was performed.

\section{Observations and Data Analysis}

   \begin{figure*}[htb]
%   \centering
   \centerline{
   \includegraphics[width=18.0cm]{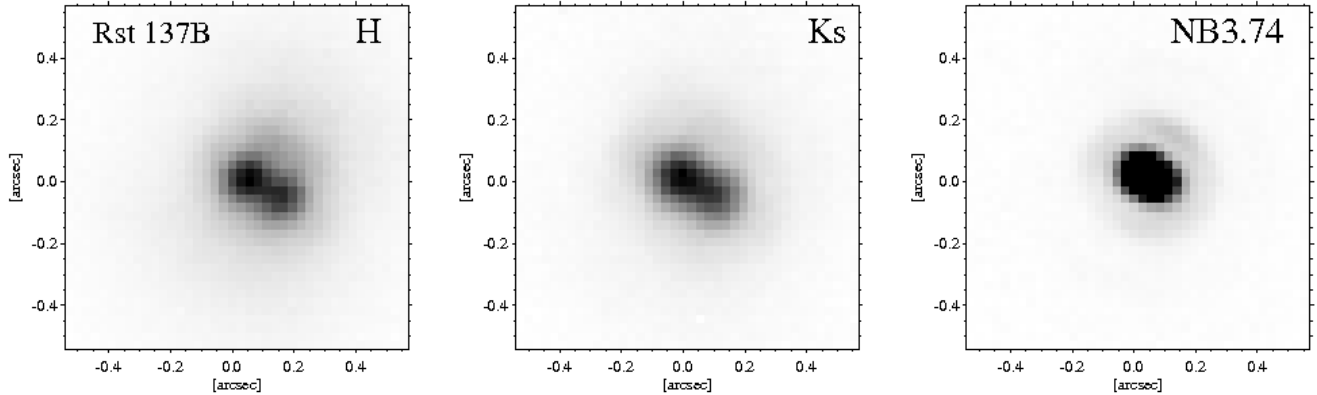}}
\caption{NACO observations of AB Dor B in H and Ks broad-band and the
NB3.74 narrow-band filter. The components of this close binary are
resolved in H and Ks. North is up, east is to the left.}
\label{naco_fig}
    \end{figure*}

Observations of AB Dor and its wide companion AB Dor B were obtained on Feb 05, 2004 with the adaptive optics instrument NACO at the ESO VLT UT4. The primary scope of the observations was a search for substellar companions, taking advantage of the newly implemented spectral Simultaneous Differential Imager (SDI, Lenzen et al. 2004). AB Dor B was also observed in direct imaging mode through broad-band H and Ks filters, and the narrow-band NB3.74 filter, and clearly resolved as a close binary (see Fig. \ref{naco_fig}). 

The binary parameters were obtained by an iterative fitting programme (see Bouy et al. 2003), using non-saturated exposures of AB Dor itself as a reference Point Spread Function (PSF).

\begin{table*}[htb]
\caption[]{Relative astrometric measurements and brightness ratios Q 
for the Ab Dor B binary. Separation (Sep) and Position Angle (P.A.) were 
derived from fits to the resolved H- and Ks-band data.}
         \label{astrometry}
\begin{tabular}{cccccc}
  Epoch & Sep. & P.A. & $Q_{\rm H}$ & $Q_{\rm Ks}$ & $Q_{\rm NB3.74}$   \\ 
        & [mas] & [deg] & & & \\ 
            \noalign{\smallskip}
            \hline
            \noalign{\smallskip}
2004.098     & 66.1$\pm$1.1 $^a$ & 238.5$\pm$1.3 & 0.79$\pm$0.01 & 0.78$\pm$0.01 & 0.79$\pm$0.02\\ 
            \noalign{\smallskip}
\end{tabular}
\begin{list}{}{}
\item[$^{\mathrm{a}}$] This corresponds to a projected separation of 0.99 AU for a distance of 14.94 pc.
\end{list}
\end{table*}

%______________________________________________________________

\section{Results and discussion}

\subsection{NACO data}

The parameters of the resolved components of AB Dor B from the NACO observations are compiled in Table \ref{astrometry}. The position angle and separation are based on a weighted mean of the parameters from each H- and Ks-band image where the weights were determined by the signal-to-noise ratio of each binary fit. Even though the H-band image is better resolved than the Ks-band image, the considerably worse Strehl ratio at H-band leads to a somewhat larger error in the brightness ratio $Q$ than at Ks-band. At NB3.74, the binary is not entirely resolved. Thus for this case, we use the separation and position angle from H- and Ks-band as fixed parameters, and fit only for $Q_{\rm NB374}$. A residual-weighted mean and error for $Q_{\rm NB374}$ was acquired by using the mean and extreme values respectively of the H- and Ks-band separation and position angle as input.

Note that the Ks-band brightness ratio given here is different from the one published in C05. This is likely due to the fact that $Q$ does not always fully converge towards minimum residuals in the Bouy et al. (2003) code, if the input (first guess) $Q_{\rm in}$ is too far from the actual $Q$. In this paper, for each image to be fitted, we have first tested a range of input values, and then manually fitted the image with a parameter range around the best-fit solution given by the automatic procedure, to ensure that it is indeed the actual minimum-residual solution. This mitigates the systematic errors of the Bouy et al. (2003) code, but it might be the case that some systematic errors remain due to errors in our reference PSF with regards to the 'true' PSF, in addition to the error bars given here.

\subsection{Isochronal age}

Due to the larger displacement of M-stars from the ZAMS than of higher-mass stars, and the better known age-luminosity relationship than for lower-mass objects, the AB Dor B binary components constitute the best candidates for isochronally dating the whole AB Dor system, especially with regards to theoretical isochrones. L05 did this by comparing AB Dor Ba and Bb to a sample of stars from the Pleiades in a $V - Ks$ versus $M_{\rm Ks}$ diagram. Since $V$ is only measured for the unresolved binary, L05 inferred individual V-band magnitudes by assuming coevality and the same relation between $\Delta V$ and $\Delta Ks$ as that of the empirical isochrone defined by the Pleiades sequence. While this analysis is in principle sound, a weakness of is obviously that it makes assumptions about a quantity that is not actually measured. In addition, the K-band brightness ratio in L05 is based on the value quoted in C05, which as we have mentioned is not the minimum-residual solution. Since we have presented here the brightness ratio also in H-band, we can improve on the analysis in L05. We use unresolved photometry in H- and Ks-band from 2MASS along with our brightness ratios to get individual H- and Ks-band magnitudes. We then compare the results to a Pleiades sample in a $H - Ks$ versus $M_{\rm Ks}$ diagram. Aside from the fact that all quantities are measured, there are several advantages of using colours $H - Ks$ over $V - Ks$: Both unresolved quantities were measured in the same survey, at almost the same time, and in addition, we can reliably compare the result with theoretical isochrones as well as empirical ones. While the models of Baraffe et al. (1998, commonly abbreviated as BCAH 98) give V-band magnitudes, they are known to correspond poorly to the actual values of these kinds of objects, whereas H- and Ks-band magnitudes are expected to be better suited (see e.g. Allard et al., 1997).

\begin{figure}[htb]
   \centering
   \includegraphics[width=8.5cm]{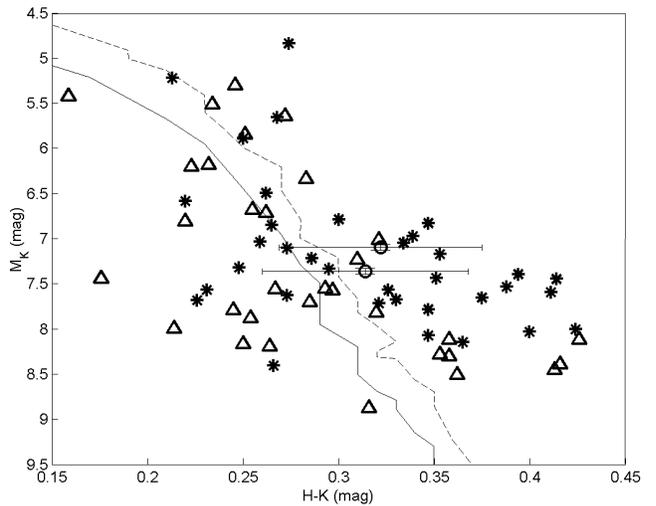}
\caption{Color-magnitude diagram with AB Dor Ba (upper circle) and AB Dor Bb (lower circle), a Pleiades sample from Steele \& Jameson, 1995 (triangles), and an IC2391 sample from Barrado y Navascu\'es, 2004 (stars). Two isochrones from BCAH 98 are also plotted -- 50 Myr (dashed line), and 100 Myr (solid line).}
\label{plei_mod}
    \end{figure}

To represent the Pleiades, we use a sample of 33 M-type stars from Steele \& Jameson (1995). The reason for this choice of sample is that well-constrained spectral types are given for all these stars in Steele \& Jameson (1995), which is useful for further analysis as we will see below. 2MASS provides H- and Ks-band photometry for each of the targets. In addition, we include a low-mass sample from Barrado y Navascu\'es (2004) to represent the young ($\sim 50$ Myr) cluster IC2391, which is also used as an age reference in L05. A plot of these samples, AB Dor Ba, AB Dor Bb and the 50 Myr and 100 Myr isochrones from BCAH 98 are shown in Fig. \ref{plei_mod}. AB Dor Ba and Bb appear to be closer in age to IC2391 than to the Pleiades, though the scatters are large, and they are closer to the 50 Myr than the 100 Myr isochrone, though of course a whole range of ages is possible within the error bars, including equal age to the Pleiades. The theoretical isochrones do not seem to fit the lower-mass end of the empirical samples, but this is to be expected. Though AB Dor Bb appears to be bluer than AB Dor Ba, in contrast with expectation from the isochrones, this is well within the error bars, and so the assumption of coevality within the AB Dor B system appears to hold.

The sizes of the error bars and large scatter of the cluster samples evidently lead to an age that is poorly constrained. The reason that the errors are so large is that the brightness difference between $H$ and $Ks$ for low-mass stars is small compared to the photometric accuracy. Unresolved photometry in more separate wavelength bands would improve the accuracy of the result. Another factor that contributes to the uncertainty is the fact that the theoretical isochrones in $H - Ks$ are somewhat jagged, which creates a small uncertainty space around each age (as is clearly evident in Fig. \ref{plei_mod}). Also, the colors of K-stars in the Pleiades have been shown to be affected by effects such as rotation and activity (see Stauffer et al., 2003). While this issue has not been studied for M-stars, it is plausible that similar effects take place in that domain. Clearly, it would be favourable to introduce additional constraints. Thus, we use the spectral type as another measured quantity to further constrain the age of the AB Dor system. Since AB Dor Ba is brighter than AB Dor Bb, we assume that its spectral type corresponds closely to the spectral type of the unresolved AB Dor B. In Vilhu et al. (1991), a spectral type range for AB Dor B of M3 to M5 is given, but Martin \& Brandner (1995) give a more well-constrained spectral type of M5 with errors of half a spectral type, and so we use the latter. Adopting this spectral type for AB Dor Ba, and using the same temperature scale as in Kenyon \& Hartman (1995), we get a temperature range of 3145 to 3305 K (see the discussion related to this in the next paragraph). By translating the known Ks-band absolute magnitude of AB Dor Ba into a bolometric magnitude using the bolometric corrections for different spectral types given in Walkowicz et al. (2004) and matching to the temperature, we can constrain the age to a range where these quantities overlap. This gives an age range of $\log(t) = 7.5$ to almost 8.0, where $t$ is the age in years, i.e. about 30 to 100 Myrs. The corresponding mass range is 0.13 to 0.2 $M_{\rm sun}$. Assuming coevality and using the known Ks-band magnitude for AB Dor Bb, we get a mass range of 0.11 to 0.18 $M_{\rm sun}$ and a temperature range of 3080 to 3240 K, corresponding to spectral types M5 to M6. Note that it follows from this reasoning that the maximal sum of the component masses is 0.38 $M_{\rm sun}$. This is consistent with the upper bound of 0.4 $M_{\rm sun}$ given in Guirado et al. (2006). Once the astrometry of all components of the AB Dor system initiated by that paper has proceeded far enough that the masses of all individual components are known, we can constrain the age even further. In this context, note that the astrometry point of the AB Dor Ba/Bb system in the Guirado et al. (2006) paper actually refers to this paper, and that the astrometric fit has been improved since then (i.e., the values quoted here should replace the corresponding values in the Guirado et al. (2006) paper).

\begin{figure*}[htb]
   \centering
   \includegraphics[width=16 cm]{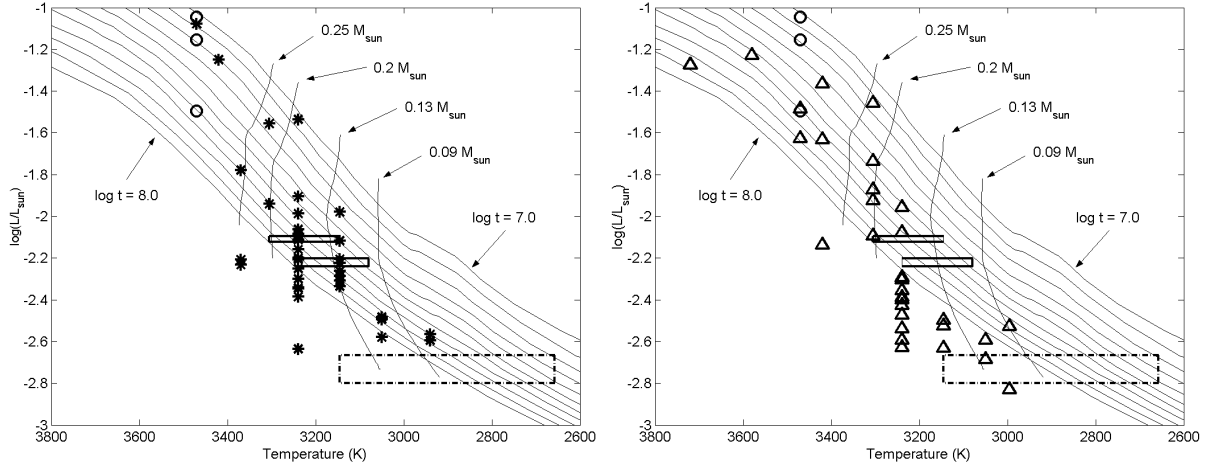}
\caption{Luminosity versus temperature for AB Dor Ba and AB Dor Bb (solid boxes). and the three M3-stars from the AB Dor moving group (circles). Left: Comparison with the IC2391 sample (stars). Right: Comparison with the Pleiades sample (triangles). Also plotted in both panels are mass tracks (units of $M_{\rm sun}$) and isochrones ($\log(t)$ changes by 0.1 per isochrone, $t$ is in units of yr) from BCAH 98. For reference, AB Dor C is also plotted (dash-dotted box), but recall that the BCAH 98 models are not expected to apply to that kind of object. Note that the temperature of AB Dor Bb is not a measured quantity, but inferred from the assumed coevality with AB Dor Ba.}
\label{bcah_mod}
    \end{figure*}

A plot corresponding to the above reasoning is shown in Fig. \ref{bcah_mod}. In addition, we include three additional stars (spectral type M3) from the AB Dor moving group (Zuckerman et al., 2004), as well as our empirical cluster samples in the same figure, after using the same procedure for finding the temperatures and bolometric luminosities. Comparing the samples, we see that AB Dor seems to be closer to the age of the IC2391 sample than the Pleiades sample, which is consistent with (and more secure than) the result of the color-magnitude diagram analysis. We also see that around the temperature range of AB Dor Ba and Bb, the means of the cluster samples seem to correspond fairly well to the expected isochrones -- about 50-60 Myr for IC2391, and more than 100 Myr for the Pleiades. Note that the earlier-type stars seem to be significantly over-luminous, which is quite unexpected. The relative positions of the three groups in the diagram are however consistent for the entire range. As mentioned, we used the temperature scale of Kenyon \& Hartman (1995). If we instead use e.g. the temperature scale of Leggett et al. (1996), the results are significantly different, since the Leggett et al. (1996) models predict a lower temperature for a given spectral type. Note that adopting such a temperature scale would lead to a significantly worse fit to the isochrones everywhere. In any case, regardless of what temperature scale is used, the relative positions of AB Dor, IC2391 and the Pleiades are the same, since the same scaling has been used for all the targets.

\begin{figure}[htb]
   \centering
   \includegraphics[width=8.5cm]{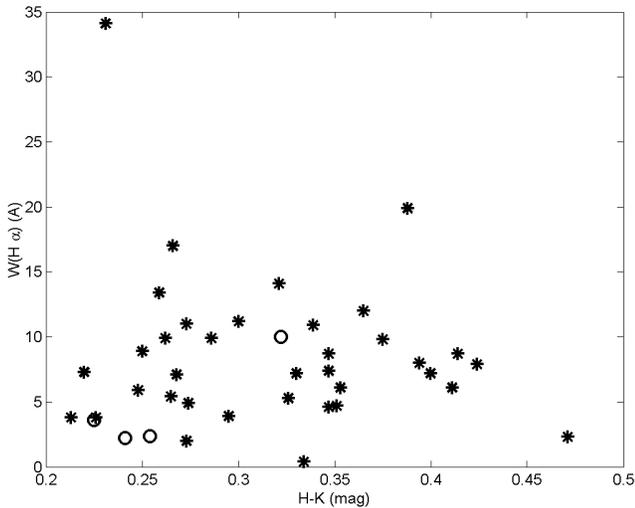}
\caption{Equivalent width of the H$\alpha$ emission of AB Dor B (top circle), three M3-stars from the AB Dor moving group (circles), and the IC2391 sample (stars). Emission is represented by positive quantites in this case.}
\label{h_alpha}
    \end{figure}

Finally, we plot the equivalent width of H$\alpha$ against the $H-Ks$ color for AB Dor B, the three M3-stars in the AB Dor moving group, and the IC2391 sample in Fig. \ref{h_alpha}. L05 do this analysis for the Pleiades, and conclude that the AB Dor moving group can not be distinguished from the Pleiades in this regard. We find that the same can be said with respect to IC2391 -- while the M3-stars of the AB Dor moving group do seem to lie near the lower edge of the IC2391 sample, AB Dor B itself is rather towards the upper edge. Hence, the overall view that AB Dor has a similar or somewhat older age than IC2391 holds also for this case.

In summary, our analysis indicates that the age of AB Dor lies between that of IC2391 and the Pleiades. As an upper limit on the age, we set the isochronal age of close to 100 Myr based on the Kenyon \& Hartman (1995) temperature scale (if the Leggett et al. (1996) scale would be used, this upper limit would be lower). A meaningful lower limit can not be set by the theoretical isochrones due to the temperature scale issue, but since we have shown that AB Dor is not younger than IC2391, we set the lower age limit of AB Dor to be the same as the age of IC2391, which according to Barrado y Navascu\'es (2004) is about 50 Myr. Hence, we end up with an age range of 50 to 100 Myr for AB Dor.

\begin{table*}[htb]
\caption[]{Summary of photometry and derived physical parameters for AB Dor Ba and Bb.}
         \label{photometry}
\begin{tabular}{cccccc}
  Component & H & K & SpT & age & mass   \\ 
        & [mag] & [mag] & & [Myr] & [$M_{\rm sun}$] \\ 
            \noalign{\smallskip}
            \hline
            \noalign{\smallskip}
Ba & 8.29$\pm$0.04 & 7.97$\pm$0.03 & M5 & 32-79 & 0.13-0.2 \\
Bb & 8.55$\pm$0.04 & 8.23$\pm$0.03 & M5-M6 & 32-79 & 0.11-0.18 \\
            \noalign{\smallskip}
\end{tabular}
\end{table*}

\subsection{Kinematical group membership}

L05 point out that the space motion of the Pleiades and AB Dor are remarkably similar -- AB Dor and its moving group are among the $\sim 0.3$\% of the stars in the Nordstr\"om et al. (2004) catalog that are closest to the mean space motion of the Pleiades. From this, L05 draw the conclusion that the AB Dor moving group and the Pleiades were part of the same star formation event, and that they should therefore be roughly coeval. However, the relative spatial positions of AB Dor and the Pleiades are of course also of importance, since if two objects share a common spatial motion, but are too far separated to have originated from the same molecular cloud, their dynamical similarities are coincidental, or at the very least insufficient for assuming coevality. From the galactic coordinates (given by SIMBAD) and using the same distances as L05, we calculate the spatial positions relative to the sun and find that the Pleiades and AB Dor are separated by about 146 pc, whereas a giant molecular cloud is typically 50 pc across. However, if the small differential velocity vector points in the right direction, the objects might converge backwards in time to a common origin at a time corresponding to the birth of the Pleiades. To check whether this is the case, we use the same galactic space motion as L05 ($U = -7.7 \pm 0.4$ km/s, $V = -26.0 \pm 0.4$ km/s and $W = -13.6 \pm 0.3$ km/s for AB Dor and $U = -6.6 \pm 0.4$ km/s, $V = -27.6 \pm 0.3$ km/s and $W = -14.5 \pm 0.3$ km/s for the Pleiades), and calculate the positions at any given time from the present epoch positions, assuming constant velocities. We find that in fact, the objects diverge backwards in time from 30 Myrs ago such that at 125 Myrs ago, the separation was $260 \pm 53$ pc. In other words, if we assume the giant molecular cloud from which both objects hypothetically formed to be 50 pc across, we can seemingly exclude the hypothesis that they did form in the same cloud. However, note again that this assumes constant velocities over $\sim 100$ Myrs, which can by no means be guaranteed to have been the case (though note that if velocities are allowed to change over time, nothing can be strictly said about a common origin based on kinematics altogether). 

In order to be more robust against changing velocities, it may be argued that using the AB Dor moving group as a whole rather than the AB Dor system alone is more relevant in the above argument. In this way, random accelerations of single objects are canceled out in a big enough sample with a common average motion. On the other hand, such an approach is risky, because misidentification of objects within the group may seriously affect the outcome. In particular, if AB Dor itself should happen not to be part of the AB Dor moving group, such an analysis would be completely irrelevant. Still, we perform this analysis in the same way as described above, adopting the same AB Dor moving group members as given in L\'opez-Santiago et al. (2006), with the same galactic space velocities. The result is that the distance between the center of the AB Dor moving group and the Pleiades is 129 pc today, and was $180 \pm 58$ pc at 125 Myrs ago. Thus the hypothesis that the two groups have separate origin is less secure in this case, but still the most plausible conclusion.

A surprising outcome of the analysis of space motion of AB Dor moving group members is that the differential motion of the individual systems with respect to the mean motion is in fact seemingly randomly distributed, rather than diverging from the center of the group as would be expected. This means that for instance, the mean separation between nearest neighbours is 17 pc at present, but would have been about 83 pc at 50 Myrs ago assuming constant velocities. Possible reasons for this behaviour are beyond the scope of this paper; we simply note that if we trust the AB Dor moving group members to have a common origin, this speaks against using a single object (AB Dor) in a differential motion analysis, as in our first example.

The inevitable conclusion of this reasoning is nonetheless that, in particular for objects of larger separation than about 50 pc at present, common origin based on common motion is insufficient by itself, but has to be coupled with other criteria such as H${\alpha}$ emission or other common age indicators. In this context, we note that in contrast to L05, L\'opez-Santiago et al. (2006) confirm the conclusion of Zuckerman et al. (2004) that the AB Dor moving group is a distinct group with an age of ~50 Myr, rather than being directly associated with the Pleiades (though Fig. 3 in their paper does seem to suggest that coevality could be possible).

   \begin{figure}[htb]
   \centering
   \includegraphics[width=8.5cm]{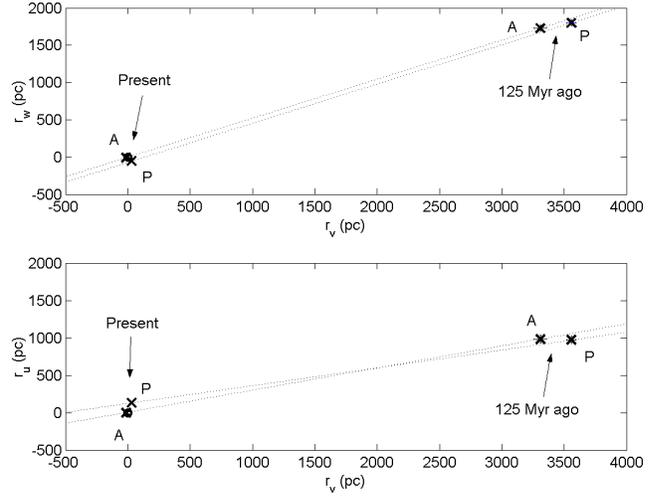}
\caption{Motions of AB Dor and the Pleiades (mean motion) assuming constant velocities. AB Dor is denoted by 'A', and the Pleiades are denoted by 'P'. $r_{\rm u}$, $r_{\rm v}$ and $r_{\rm w}$ are the coordinates corresponding to $U$, $V$ and $W$. The dotted lines are the spatial tracks relative to the sun over time. The circle at origin is the sun.}
\label{sp_mot}
    \end{figure}

\subsection{Consequences for AB Dor C}

\begin{figure}[htb]
   \centering
   \includegraphics[width=8.5cm]{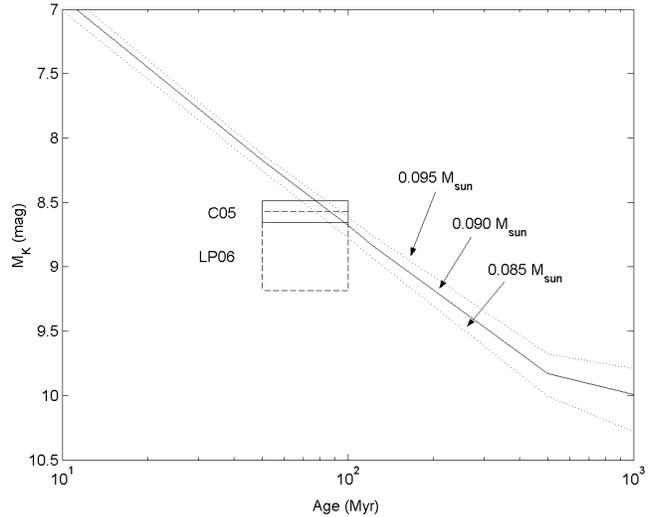}
\caption{Comparison of the measured and estimated quantities of AB Dor C to the Chabrier et al. (2000) model predictions. While there is an overlap between the measured brightness and the mass tracks for the given age, most of the parameter space implies that AB Dor C is fainter than what is predicted by the models. The solid box corresponds to the $M_{\rm Ks}$ measurements in C05, and the dashed box to Luhman \& Potter (2006).}
\label{chab_mod}
    \end{figure}

For comparing AB Dor C with theoretical models, we adopt the age range of 50 to 100 Myrs, and check independently the Ks-band absolute magnitude ranges given by C05 and Luhman \& Potter (2006), i.e. $9.45^{+0.06}_{-0.075}$ mag and $9.79^{+0.25}_{-0.33}$ mag, respectively. For easy comparison with previous work, we use the evolutionary model of Chabrier et al. (2000). The result can be seen in Fig. \ref{chab_mod}. It is easily seen that while the C05 brightness range implies a possible overlap with the models for the higher part of the age range, only a very small part of the Luhman \& Potter (2006) range is consistent with the models. In total, a minority of the parameter space is consistent with the models.

\section{Conclusions}

The age of the AB Dor system is fundamental for using AB Dor C as a calibration point for theoretical evolutionary models for young, low-mass stars. L05 use comparison of AB Dor B with empirical isochrones and kinematic analysis to determine that AB Dor is roughly coeval with the Pleiades, and has an age of 75 to 150 Myrs. With such an age range, AB Dor C has a mass-luminosity relationship which is consistent to what is predicted by the Chabrier et al. (2000) model. We conclude that the kinematic similarities between the Pleiades and AB Dor are insufficient to infer a common origin, and our comparisons with both theoretical and empirical isochrones imply a younger age range. A detailed comparison gives ages of 50 to 100 Myrs. Applying this age range to the case of AB Dor C, we find that while consistency with theoretical models is possible, the object more likely has a lower brightness than predicted by the models, if it is a single object. It has on the other hand been suggested that AB Dor C could be an unresolved binary (see Marois et al., 2005). If this turns out to be the case, AB Dor C may still be entirely consistent with the models.

The case of AB Dor and its role as a data point for calibrating the mass-luminosity relationship as a function of age for young, low-mass stars can be further constrained with additional measurements. Future work will include spatially resolved spectroscopy of the AB Dor B binary in order to better constrain the individual spectral types, and monitoring of the orbital motion in order to constrain the individual masses. Assuming a semi-major axis of $\sim$1 AU and a system mass of $\sim$0.375 $M_{\rm sun}$, the orbital period is $\sim$1.6 years, meaning that the orbital parameters should be readily measurable within a relatively short time-frame. Constraints can also be made with regards to AB Dor C -- the photometry of AB Dor C can likely be improved, and radial velocity measurements of this component could be taken in order to determine whether or not it is in fact an unresolved binary.

\begin{acknowledgements}
We acknowledge support by the Bundesministerium f\"ur Wirtschaft und Technologie through the Deutsche Zentrum f\"ur Luft- und Raumfahrt (F\"orderkennzeichen 50 OR 0401). We thank Ben Zuckerman, Stan Metchev, and an anonymous referee for helpful comments.
\end{acknowledgements}

\end{document}